\documentclass[aps,twocolumn,prl,superscriptaddress]{revtex4-2}
\usepackage[tbtags,fleqn]{amsmath}
\usepackage{amssymb}
\usepackage{amsfonts}
\usepackage{mathtools}
\usepackage[colorlinks=true,citecolor=blue,urlcolor=blue]{hyperref}
\hypersetup{colorlinks,linkcolor=blue,filecolor=blue,urlcolor=blue,citecolor=blue}
\usepackage{multirow}
\usepackage{graphicx}
\usepackage{subfigure}
\usepackage{cases}
\usepackage[normalem]{ulem}
\usepackage{xr}

\usepackage{appendix}
\usepackage{tabularx} 

\usepackage{times}

\usepackage{xr}
\begin{document}
\title{Unified Framework for Quantum Resource Recycling via Instrument-Dependent Back-action
}
	
\author{Zinuo Cai}
\affiliation{Key Laboratory of Low-Dimension Quantum Structures and Quantum Control of Ministry of Education, Synergetic Innovation Center for Quantum Effects and Applications, Department of Physics, Hunan Normal University, Changsha 410081, China}
\affiliation{Hunan Research Center of the Basic Discipline for Quantum Effects and Quantum Technologies, Hunan Normal University, Changsha 410081, China}
\affiliation{Institute of Interdisciplinary Studies, Hunan Normal University, Changsha 410081, China}	
	
\author{Jianxin Song}
\affiliation{Key Laboratory of Low-Dimension Quantum Structures and Quantum Control of Ministry of Education, Synergetic Innovation Center for Quantum Effects and Applications, Department of Physics, Hunan Normal University, Changsha 410081, China}
\affiliation{Hunan Research Center of the Basic Discipline for Quantum Effects and Quantum Technologies, Hunan Normal University, Changsha 410081, China}
\affiliation{Institute of Interdisciplinary Studies, Hunan Normal University, Changsha 410081, China}
\author{Ali Asadian}
\affiliation{Physics Department, Institute for Advanced Studies in Basic Sciences (IASBS), Gava Zang, P.O. Box 45137-66731, Zanjan, Iran}
\author{Changliang Ren}\thanks{Corresponding author: renchangliang@hunnu.edu.cn}
\affiliation{Key Laboratory of Low-Dimension Quantum Structures and Quantum Control of Ministry of Education, Synergetic Innovation Center for Quantum Effects and Applications, Department of Physics, Hunan Normal University, Changsha 410081, China}
\affiliation{Hunan Research Center of the Basic Discipline for Quantum Effects and Quantum Technologies, Hunan Normal University, Changsha 410081, China}
\affiliation{Institute of Interdisciplinary Studies, Hunan Normal University, Changsha 410081, China}
	
\begin{abstract}

{Accurately characterizing measurement backaction is essential to understanding the limits of quantum correlation sharing in sequential scenarios. Here, we develop a unified quantum-instrument framework that attributes correlation sharing to backaction governed by the Kraus decomposition, thereby unifying seemingly distinct strategies such as weak measurements and probabilistic projective measurements (PPMs). This framework further enables us to derive more general analytical conditions for sharing Bell nonlocality among arbitrarily many sequential observers in unilateral scenarios. Moreover, we demonstrate the possibility of bilateral fourfold Bell-nonlocality sharing under unbiased measurement settings, a result previously widely regarded as impossible. This phenomenon arises from the interplay between the freedom in Kraus decompositions, which enables outcome-conditioned local unitary feedback, and the feed-forward of previous measurement outcomes to subsequent observers, which allows their measurement settings to adapt accordingly. Together, these ingredients enable four pairs of observers to simultaneously violate the CHSH inequality, highlighting the crucial role of quantum instruments beyond the statistics captured by positive-operator-valued measures (POVMs) alone.
}
\end{abstract}


	
\maketitle

\textit{Introduction.---}
Quantum correlations, such as entanglement and Bell nonlocality, constitute fundamental resources for quantum information processing~\cite{Brunner.RevModPhys.86.419,horodecki2009quantum,Uola.RevModPhys.92.015001,Tavakoli.Rep.Prog.Phys.85.056001_2022,Cavalcanti_2017}. However, due to the unavoidable backaction induced by measurements, these correlations are typically degraded during information extraction and are often regarded as “single-use” resources~\cite{Busch_1996_quantum,Aharonov_2005_Quantum_Paradoxes,Nielsen2010,wheeler_2014_quantum}. The trade-off between information extraction and measurement-induced disturbance is therefore a central challenge for enabling the repeated use of quantum correlations in sequential multi-observer scenarios~\cite{Aharonov_1964_PhysRev.134.B1410,Aharonov_1988_PhysRevLett.60.1351,Abraham_2012_Nonperturbative,nogues_1999nature_seeing,Pryde_2004_PhysRevLett.92.190402,gudder_2001_sequential,Sciarrino_2006_PhysRevLett.96.020408,Radim_2011_PhysRevA.83.032311,Asadian_Covariant_2026}. In 2015, Silva~\textit{et al.}~\cite{Silva.Ralph_PhysRevLett.114.250401_2015} investigated the sharing of Bell nonlocality among multiple sequential observers, revealing a key mechanism by which quantum measurements can enable resource reuse. This work sparked broad interest in reusable quantum correlations~\cite{cairen_2024_review,Ren.Changliang_PhysRevA.100.052121_2019,Ren.Changliang_PhysRevA.105.052221_2022,YaoDan.PhysRevA.103.052207_2021,Bera.Anindita_PhysRevA.98.062304_2018,Hou.Wenlin_PhysRevA.105.042436_2022}, whose sustainable utilization is becoming increasingly important for scalable and networked quantum technologies~\cite{Wang.Jian.Hui_PhysRevA.106.052412_2022,Cai_jpafull_2024,Maity.Ananda.G_PhysRevA.101.042340_2020,
Karthik.Mohan_NEW.J.PHYS.10.1088.1367-2630.ab3773_2019,
Miklin.Nikolai_PhysRevResearch.2.033014_2020,
Chirag_PhysRevA.103.032408_2021,
Curchod.F.J_PhysRevA.95.020102_2017,
Xue-Bi.An_Opt.Lett.ol-43-14-3437_2018}.

Several strategies for sequential correlation sharing, ranging from weak measurements~\cite{Silva.Ralph_PhysRevLett.114.250401_2015,Ren.Changliang_PhysRevA.100.052121_2019,Ren.Changliang_PhysRevA.105.052221_2022,YaoDan.PhysRevA.103.052207_2021} to asymmetric POVMs~\cite{Brown.Peter.J_PhysRevLett.125.090401_2020,Zhang.Tinggui_PhysRevA.103.032216_2021,Cai_pra_entanglement} and probabilistic projective measurements (PPMs)~\cite{Anna._PhysRevLett.129.230402_2022,Sasmal_PhysRevLett.133.170201_2024,dong_2024_sharing,zhangtg_PhysRevA.109.022419_2024}, have been proposed. Yet their underlying physical mechanisms remain disconnected, preventing systematic comparison and obscuring the origin of their differences. The fragmented treatment makes it unclear whether the observed differences between schemes arise from measurement statistics or from deeper backaction mechanisms. Consequently, critical questions remain unresolved: Is bilateral sharing of Bell nonlocality strictly forbidden under unbiased measurement choices, as conventionally asserted~\cite{Cheng.Shuming_PhysRevA.104.L060201_2021,Zhu.Jie_PhysRevA.105.032211_2022}? Furthermore, within a general measurement formalism, what physical constraints dictate the ultimate limits of unbounded unilateral sharing~\cite{Brown.Peter.J_PhysRevLett.125.090401_2020,Sasmal_PhysRevLett.133.170201_2024}?

In this letter, we revisit the fundamental boundaries of quantum correlation sharing from the perspective of quantum instruments rather than measurement statistics. By developing a Kraus-decomposition-based functional framework~\cite{kraus1971general,kraus1983states}, we unify seemingly distinct sequential measurement strategies, including weak measurements and PPM protocols, within a single formalism. In this picture, different sharing models arise not from distinct physical mechanisms, but from different quantum instruments associated with the same POVMs. The framework thus reveals that their essential differences originate from measurement backaction rather than outcome statistics. Notably, this formulation is fully general and encompasses all previously proposed sequential sharing schemes without additional assumptions, providing a unified and transparent description.

Based on this unified framework, we resolve the above open questions. 
{\color{black}
Firstly, exact bounds for infinite sequential sharing of Bell nonlocality in unilateral scenarios are provided under different Kraus decompositions. Secondly,  we demonstrated that bilateral Bell nonlocality sharing is achievable under unbiased measurements.}
These results show that the ultimate boundaries of quantum correlation sharing are determined not only by POVMs but also by the underlying quantum instrument structure. Exploiting the freedom in Kraus decompositions reveals new possibilities and provides a meaningful route to deeply understanding the fundamental properties of correlation sharing.

\begin{figure}[t!]
\centering  
\includegraphics[width=0.45\textwidth]{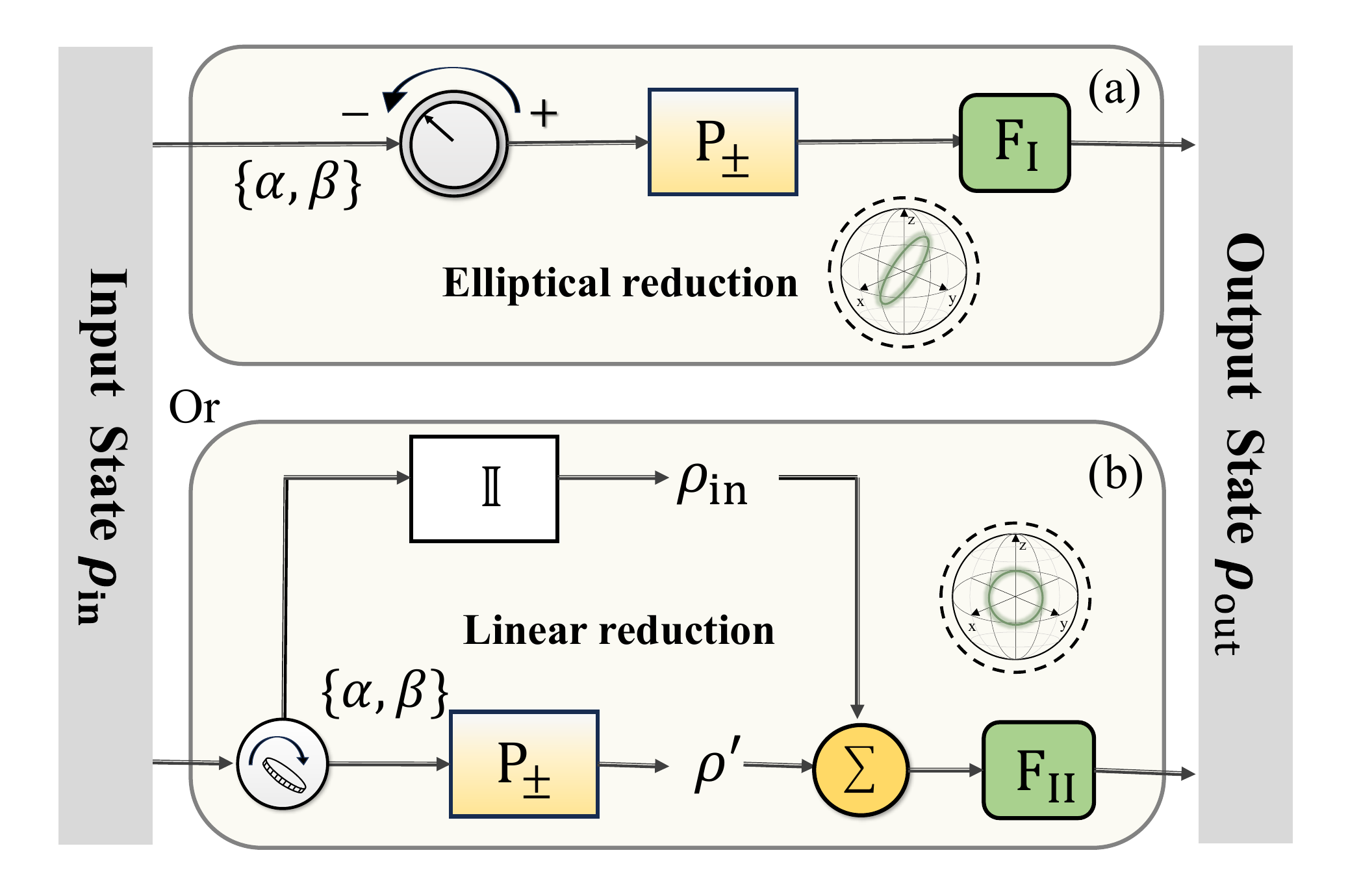}
\caption{\small{
Instrument-dependent measurement backaction. (a) Elliptical reduction model. Coherent superposition of measurement channels (Lüders decomposition) constrains the transverse coherence reduction factor $F_{\mathrm{I}}$ to an elliptical trajectory dictated by sharpness $\alpha$ and bias $\beta$. (b) Linear reduction model. An incoherent probabilistic mixture ($\Sigma$) of completely dephasing ($P_{\pm}$) and non-disturbing ($\mathbb{I}$) branches yields a linear scaling $F_{\mathrm{II}}$. The elliptical mode strictly provides greater backaction suppression ($F_{\mathrm{I}} \ge F_{\mathrm{II}}$).
}}
\label{fig:my_image}
\end{figure}


\textit{
Kraus Decomposition-Dependent measurment Backaction and  Correlation Sharing---}
Quantum measurement yields information at the cost of inevitable backaction, which drives state evolution and the dissipation of resources such as coherence~\cite{coherence_RevModPhys.89.041003}. Although information gain and disturbance arise simultaneously, measurement statistics and conditional state evolution are formally decoupled: the former is determined solely by the POVM $\{E_\pm\}$~\cite{wiseman2009quantum}, while the latter depends on the underlying quantum instrument through the specific Kraus operators $\{K_\pm\}$, yielding post-measurement states $\rho_\pm = K_\pm \rho K_\pm^\dagger / \mathrm{Tr}(\rho E_\pm)$. The non-uniqueness of the Kraus decomposition $E_\pm = K_\pm^\dagger K_\pm$~\cite{kraus1971general,kraus1983states} implies that identical measurement statistics can stem from diverse physical implementations, each exerting a distinct dynamical backaction.

Without loss of generality, an arbitrary single-qubit binary measurement observable can be defined as $M = \alpha \vec{n} \cdot \vec{\sigma} + \beta\mathbb{I}$,
where $\alpha \in (0,1]$ and $\beta \in (-1,1)$ represent the measurement sharpness and outcome bias, respectively, with the positivity constraint $\alpha + |\beta| \le 1$. The corresponding POVM elements take the form
\begin{equation}\label{eq:POVM_def}
E_\pm = \frac{1 + \alpha \pm \beta}{2} P_\pm
      + \frac{1 - \alpha \pm \beta}{2} P_\mp,
\end{equation}
where the projectors are defined as $P_\pm = \frac{\mathbb{I} \pm \vec{n}\cdot\vec{\sigma}}{2}$. For any Kraus decomposition $\{K_\pm\}$ realizing this POVM, the post-measurement state is given by
\begin{equation}
\mathcal{E}(\rho) = \sum_{i=\pm} K_i \rho K_i^\dagger 
=  \sum_{i=\pm} P_i \rho P_i 
+ F \sum_{i\neq j} P_i \rho P_j,
\end{equation}
where the reduction factor $F$ depends on the specific Kraus decomposition. This factor quantifies the measurement backaction by characterizing the rescaling of the off-diagonal elements of the post-measurement state relative to those of the initial state in the projective measurement basis $\{P_\pm\}$. The detailed derivation is presented in the Appendix. 

Two typical Kraus decompositions are considered. The first corresponds to the conditional evolution governed by the canonical square-root Kraus operators, known as the Lüders decomposition:
\begin{equation}
K_\pm = \sqrt{E_\pm}=\sqrt{\frac{1 + \alpha \pm \beta}{2}} P_\pm + \sqrt{\frac{1 - \alpha \pm \beta}{2}} P_\mp.
\end{equation}
In this case, the retained coherence originates from the coherent superposition of measurement channels, leading to the reduction factor $F_{\mathrm{I}} =\frac{\sqrt{(1-\alpha-\beta)(1+\alpha-\beta)}+ \sqrt{(1-\alpha+\beta)(1+\alpha+\beta)}}{2}$. It can give an intuitive geometric constraint
$\frac{\alpha^2}{1-F_{\mathrm{I}}^2}
+\frac{\beta^2}{F_{\mathrm{I}}^2}=1$,
which explicitly characterizes the bounds imposed by measurement backaction. In the unbiased-outcome limit ($\beta=0$), the relation reduces to
$F_{\mathrm{I}}=\sqrt{1-\alpha^2}$,
corresponding to a circular coherence-retention trajectory. For ($\beta \neq 0$), the rotational symmetry is broken and an absolute bound $F_{\mathrm{I}} \ge |\beta|$ emerges, deforming the trajectory into an ellipse [Fig.~\ref{fig:my_image}(a)]. Owing to this distinctive geometric structure, this evolution can be referred to as the \textit{elliptical reduction mode}. In contrast, the second evolution mode replaces the coherent superposition of measurement channels with a mixed-channel decomposition, effectively reducing the measurement to a classical probabilistic mixture [Fig.~\ref{fig:my_image}(b)]. Formally, the POVM elements $E_\pm = \sum_{i=1}^{2}K_{\pm}^{(i)\dagger}K_{\pm}^{(i)}$ are realized through Kraus operators that decouple into two independent physical branches: a sharp projective branch that induces complete decoherence, and an identity branch that preserves the pre-measurement state,
\begin{equation}\label{kraus-2}
	K_{\pm}^{(1)} = \sqrt{\alpha} P_\pm, \qquad K_{\pm}^{(2)} =\sqrt{\frac{1 - \alpha \pm \beta}{2}} \mathbb{I}.
\end{equation}
Since the residual coherence stems exclusively from the non-disturbing identity branch, the total reduction factor exhibits a strictly linear scaling, $F_{\mathrm{II}} = 1-\alpha$, defining the \textit{linear reduction mode}.  It is worth noting that the analytical gap between these two modes is captured by the global bound $F_{\mathrm{I}} \ge F_{\mathrm{II}}$ [Fig.~\ref{fig:my_image}], confirming that the elliptical mode suppresses backaction more effectively than its linear counterpart. This instrument-dictated disparity is of fundamental consequence in sequential measurement scenarios, where a quantum state evolves through a cascade of independent observers. Since each subsequent measurement acts on the conditional state generated by the preceding observer, the measurement-induced backaction governs the dynamical degradation of quantum correlations, thereby determining whether quantum resources, such as entanglement and nonlocality, can be sequentially shared among multiple observers.

This framework can directly unify the principal strategies developed in sequential measurement scenarios~\cite{cairen_2024_review} and establishes a general paradigm for quantifying the recycling of quantum resources. Previously proposed strategies for sequential correlation sharing—including weak measurements, asymmetric POVMs, and PPMs—emerge as special cases within this unified framework. In the unbiased-outcome limit ($\beta=0$), the model recovers the two foundational mechanisms introduced by Silva \textit{et al.}~\cite{Silva.Ralph_PhysRevLett.114.250401_2015}: the linear reduction corresponds to the square-pointer weak measurement, while the elliptical reduction captures the optimal weak-measurement pointer. Owing to its superior efficiency in suppressing measurement backaction, the elliptical scheme has been widely adopted in subsequent sequential-sharing studies, such as that of Brown \textit{et al.}~\cite{Brown.Peter.J_PhysRevLett.125.090401_2020}. Furthermore, the PPM strategy corresponds to the linear mode with $\beta = 1-\alpha$, which exactly reproduces the sequential measurement model recently proposed by Sasmal \textit{et al.}~\cite{Sasmal_PhysRevLett.133.170201_2024}.

\begin{table*}[htbp]
	\centering 
	\caption{\small Maximum achievable CHSH values under Unilateral (Uni.) and Bilateral (Bi.) scenarios for F$_{\text{I}}$ or F$_{\text{II}}$.}
	\label{tab:comparison}
	\begin{tabular}{ccccc}
        \toprule	
		Strategy & 
		\qquad Max. (Uni.-F$_{\text{I}}$)\qquad & \qquad Max. (Uni.-F$_{\text{II}}$) \qquad & 
		\qquad Max. (Bi.-F$_{\text{I}}$)\qquad  &  \qquad Max. (Bi.-F$_{\text{II}}$) \qquad \\
		\hline 
		

		
		\multirow{1}{*}{Weak $\{\beta =0\}$} & 
		\textbf{2.2627\textsuperscript{\cite{Silva.Ralph_PhysRevLett.114.250401_2015}}} & 2.0354 & 
		None\textsuperscript{\cite{Cheng.Shuming_PhysRevA.104.L060201_2021}} & None \\
		
		\hline
		
		
		\multirow{1}{*}{ \,PPM  $\{\beta=1-\alpha\}$\,} &
		2.1378 & \textbf{2.0569}\textsuperscript{\cite{Sasmal_PhysRevLett.133.170201_2024}} & 
		\textbf{2.0042} & \textbf{2.00123} \\
		
		\hline\hline 
	\end{tabular}
	
	\vspace{1ex} 
	\parbox{0.7\textwidth}{ 
		\footnotesize
		\raggedright
		\textit{Note:} 
       Bold denotes the optimal measurement strategy for each scenario.
	}
\end{table*}

\textit{Unilateral Sharing of Bell Nonlocality---}
{\color{black}
A unilateral sequential scenario is considered in which a pure entangled two-qubit state, $|\Phi(\theta)\rangle=\cos\theta |00\rangle+\sin\theta |11\rangle$, is distributed between a single observer $A_1$ and $N$ sequential observers $B_k$ ($1\le k\le N$). Upon receiving local binary inputs $x_1,y_k\in\{0,1\}$, the observers perform generalized POVM measurements characterized by the parameter sets $\{\alpha^{(1)}_{A_{x_1}}, \beta^{(1)}_{A_{x_1}}, \vec{n}^{(1)}_{A_{x_1}}\}$ and $\{\alpha^{(k)}_{B_{y_k}}, \beta^{(k)}_{B_{y_k}}, \vec{n}^{(k)}_{B_{y_k}}\}$, respectively, as defined in Eq.~(\ref{eq:POVM_def}). Using this framework, we systematically derive the exact parameter regimes under which all $N$ observer pairs $\{A_1, B_k\}$ exhibit simultaneous Clauser-Horne-Shimony-Holt (CHSH) violations~\cite{chshPhysRevLett.23.880}.
This approach generalizes prior results restricted to standard weak measurements or PPMs~\cite{Brown.Peter.J_PhysRevLett.125.090401_2020, Sasmal_PhysRevLett.133.170201_2024}.}

\textit{Result 1.} The theoretical bounds for the infinite unilateral sharing of quantum nonlocality are analytically established. It is clearly shown that, under both the generalized elliptical and linear reduction models, there exist specific initial entangled states and measurement sequences that enable a single observer to simultaneously share nonlocality with an arbitrary number of sequential observers.
{\color{black}The theoretical bounds for the unbounded unilateral sharing of quantum nonlocality are analytically established. Rather than merely confirming the existence of such sharing, we provide a fine-grained characterization of the necessary conditions under both generalized elliptical and linear reduction models. }




In this scenario, all measurements are performed in the $x\text{-}z$ plane, where $A_1$ and each $B_k$ adopt the orientations $\{\phi, -\phi\}$ ($\phi \in (0, \pi)$) and $\{0, \pi/2\}$, respectively.
While $A_1$ and the terminal observer $B_N$ perform sharp projective measurements, the intermediate observers $B_k$ ($k < N$) employ an asymmetric measurement strategy. Specifically, their measurement parameters are set as $\{\alpha^{(k)}_{B_{y_k}}\}_{y_k \in \{0,1\}}=\{1,\alpha^{(k)}_{B}\}$ and $\{\beta^{(k)}_{B_{y_k}}\}_{y_k \in \{0,1\}}=\{0,\beta^{(k)}_{B}\}$, yielding corresponding reduction factors $\{F^{(k)}_{B_{y_k}}\}_{y_k \in \{0,1\}} = \{0, F^{(k)}_{B}\}$. 
Consequently, the CHSH value $\mathcal{S}_k$ for any observer pair $(A_1, B_k)$ can be given as
\begin{equation}
\begin{split}
\mathcal{S}_{k} = 2 \Biggl[ & \sin(2\theta) \mathcal{T}_{k-1} \cos\phi \\
& + \biggl( \Bigl(\frac12\Bigr)^{\!k-1} \alpha^{(k)}_{B_{\!y_k}} 
+ \cos(2\theta) \beta^{(k)}_{B_{\!y_k}} \biggr) \sin\phi \Biggr],
\end{split}
\end{equation}
where $\mathcal{T}_{k-1} = \prod_{i=1}^{k-1} \frac{1+F^{(i)}_{B}}{2}$ (with $\mathcal{T}_{0} \equiv 1$). 
By jointly imposing the CHSH violation condition $\mathcal{S}_k > 2$ and the positivity constraint $\alpha^{(k)}_{B} + \beta^{(k)}_{B} \le 1$ (assuming $\beta^{(k)}_{B} > 0$), the admissible measurement parameter regime achieving CHSH violations can be rigorously determined.
Specifically, within the regime $\cos 2\theta < 2^{1-k}$, the lower-bound condition
\begin{equation}\label{single-LB}
	\alpha^{(k)}_{B} > \frac{2^{k-1} \left(1 - \cos\phi \sin 2\theta \cdot \mathcal{T}_{k-1} - \cos 2\theta \sin\phi\right)}{\left(1 - 2^{k-1} \cos 2\theta\right) \sin\phi}\equiv \mathrm{L}_k,      
\end{equation}
guarantees the existence of a valid measurement parameter $\beta^{(k)}_{B}$ within the interval $\frac{1 - 2^{1-k} \alpha^{(k)}_{B} \sin\phi - \mathcal{T}_{k-1} \cos\phi \sin 2\theta}{\sin\phi \cos 2\theta} < \beta^{(k)}_{B} \leq 1 - \alpha^{(k)}_{B}$ for $\theta \neq \pi/4$. For the maximally entangled case ($\theta = \pi/4$), this requirement naturally reduces to $\beta^{(k)}_{B} \leq 1 - \alpha^{(k)}_{B}$. 
Formalizing these analytic boundaries, we present the following theorem.


\textit{Theorem 1.} For any sequence length $N \ge 2$, provided the measurement angle is chosen as $\phi = \frac{1}{c}(\frac{\pi}{2} - 2\theta)$, an initial entangled state satisfying $\cos 2\theta < 2^{1-N}$ guarantees the existence of a parameter sequence $\{\alpha^{(k)}_{B}\}_{k=1}^{N}$ such that
\begin{equation}\label{sequence}
	0<\mathcal L_k <\alpha^{(k)}_{B}<1
\end{equation}
holds for all $1 \le k \le N$. 

Here, $\mathcal L_k$ is obtained by substituting $\phi$ into the definition of $\mathrm{L}_k$ in Eq.~(\ref{single-LB}). For the elliptical reduction model, $c$ remains a finite constant (with $c=1$ in the linear case). Moreover, within the completeness constraints, the outcome-bias parameters $\{\beta^{(k)}_{B}\}_{k=1}^{N}$ can be freely chosen from their respective admissible intervals.

{\color{black}\textit{Proof.—} We construct a sequence of measurement parameters $\{\alpha^{(k)}_B\}$ strictly bounded slightly above their theoretical minima, such that $\alpha^{(k)}_B > \mathcal{L}_k$ for all $k$, as detailed in {\color{black}Sec.~II of} the Supplemental Material [SM]. To ensure positivity ($\alpha^{(k)}_B > 0$) and avert singularities, the initial state must satisfy the threshold condition $\cos 2\theta < 2^{1-N}$. By introducing the parameterized construction $\phi = \frac{1}{c}\left(\frac{\pi}{2} - 2\theta\right)$, this constraint is naturally satisfied in the small-angle limit $\phi \to 0^+$. To further satisfy the upper bound ($\alpha^{(k)}_B < 1$), we fix the bias parameters $\{\beta^{(k)}_B\}$ to finite constants within the completeness constraint, and analyze the scaling behavior of the sequence. Specifically, in the generalized elliptical reduction model ($c \in \mathbb{Z}^+$), the sequence obeys a uniform first-order scaling, $\alpha^{(k)}_B \sim \mathcal{O}(\phi)$, which smoothly vanishes as $\phi \to 0^+$. Conversely, in the linear reduction model ($c=1$), the initial lower bound degenerates to $\mathcal{L}_1=0$ and each recurrence step introduces an $\mathcal{O}(\phi^{-1})$ amplification. Setting the initial term to $\alpha^{(1)}_B \sim \mathcal{O}(\phi^{N})$ offsets this cumulative divergence across the $N$ sequential measurements, confining the entire sequence to $\alpha^{(k)}_B \lesssim \mathcal{O}(\phi)$ and ensuring it uniformly collapses to zero as $\phi \to 0^+$. Consequently, regardless of the specific reduction mechanism, selecting a sufficiently small $\phi > 0$ restricts the entire sequence $\{\alpha^{(k)}_B\}$ to the physical domain $(0, 1)$. This rigorously guarantees the existence of valid parameters for simultaneous CHSH violations across all $N$ observer pairs.}

Notably, all previously reported schemes for unilateral unbounded sharing arise as special cases of Result 1. Under the elliptical reduction model, selecting a maximally entangled initial state together with unbiased weak measurements ($\beta_B^{(k)}=0$) recovers the result of Brown \textit{et al.}~\cite{Brown.Peter.J_PhysRevLett.125.090401_2020}. In the linear reduction model, choosing $\phi=\pi/2-2\theta$ and $\beta_B^{(k)}=1-\alpha_B^{(k)}$ reduces exactly to the PPM-based scheme of Sasmal \textit{et al.}~\cite{Sasmal_PhysRevLett.133.170201_2024}. Result 1 therefore integrates all known schemes within a unified framework and, more importantly, establishes the most general conditions for unbounded unilateral nonlocality sharing.



Moreover, the effectiveness of different sequential measurement strategies is related to the reduction model. For example, in the $k=2$ sequential scenario, under the elliptical reduction model, the unbiased weak measurement strategy can achieve a maximum double violation value of 2.26 (strictly consistent with the results of Silva et al.~\cite{Silva.Ralph_PhysRevLett.114.250401_2015}), significantly higher than the theoretical upper limit of 2.138 for the PPM strategy. Conversely, in the linear reduction model, the maximum double violation value achievable by the PPM strategy is 2.057~\cite{Sasmal_PhysRevLett.133.170201_2024}, exceeding the maximal value of 2.035 for the unbiased weak measurement strategy (see Table~\ref{tab:comparison}). This result indicates that neglecting the feedback from the measurement instrument is insufficient for a comprehensive evaluation of sequential measurement strategies.

\textit{Bilateral Sharing of Bell Nonlocality---}
For a long time, sharing bipartite Bell nonlocality under unbiased measurement-setting choices was believed to be subject to stringent limitations~\cite{Cheng.Shuming_PhysRevA.104.L060201_2021,Zhu.Jie_PhysRevA.105.032211_2022}. Exploiting the nonuniqueness of Kraus decompositions, we introduce tunable unitary freedom via $\widetilde K_i=U_i(\eta)K_i$, which preserves the POVM elements, and combine it with local feed-forward of measurement outcomes. We demonstrate that this approach enables Bell nonlocality to be shared among arbitrary pairs of sequential observers while maintaining unbiased measurement-setting choices.

An initial entangled state $|\Phi(\theta)\rangle$ is distributed between spatially separated Alice and Bob, each hosting sequential observers $A_k$ and $B_k$, with $k\in\{1,2\}$. The first observers employ generalized quantum instruments characterized by  $\{\alpha^{(1)}_{A_{x_1}},\beta^{(1)}_{A_{x_1}},
\phi^{(1)}_{A_{x_1}},\eta^A_{a_1|x_1}\}$ and $\{\alpha^{(1)}_{B_{y_1}},\beta^{(1)}_{B_{y_1}},
\phi^{(1)}_{B_{y_1}},\eta^B_{b_1|y_1}\}$, respectively. After the measurements, the quantum systems are passed to the respective second observers, who use the local outcomes $a_1$ and $b_1$ to select their measurement directions  $\{\phi^{(2)}_{A_{x_2}|a_1}\}$ and $\{\phi^{(2)}_{B_{y_2}|b_1}\}$, respectively. The second observers perform sharp projective measurements. 
Although their measurement directions depend on the preceding outcomes, each observable $A_{x_2}$ ($B_{y_2}$) is constructed from the two conditional directions $\{\phi^{(2)}_{A_{x_2}|\pm}\}$($\{\phi^{(2)}_{B_{y_2}|\pm}\}$)\sout{,}{\color{black}. The measurement inputs remain independently and uniformly distributed,} thereby preserving unbiased measurement-setting choices. 
The protocol involves only local quantum-state transmission and classical feed-forward, with no communication between Alice and Bob. 
The combined effect of unitary freedom and local feed-forward can be stated as follows.

\textit{Result 2.} Analytical and numerical results demonstrate that, under unbiased measurement choices, the two-qubit entangled state can be sequentially recycled such that the measurement statistics of all four observer pairs, $(A_1,B_1)$, $(A_1,B_2)$, $(A_2,B_1)$, and $(A_2,B_2)$, simultaneously violate the CHSH inequality~\cite{chshPhysRevLett.23.880}, thereby demonstrating bilateral sharing of Bell nonlocality.

We first analytically establish the simultaneous violation of all four CHSH inequalities. We adopt the PPM strategy, $\beta=1-\alpha$, and choose the initial-observer measurement strengths $\{\alpha^{(1)}_{A_0},\alpha^{(1)}_{A_1}\}=\{\frac{11}{16},\frac{13}{18}\}$ and $\{\alpha^{(1)}_{B_0},\alpha^{(1)}_{B_1}\}=\{\frac{3}{16},\frac{20}{21}\}$, with directions $\phi^{(1)}_{A_0}=\phi=\frac{7\pi}{16}$, $\phi^{(1)}_{A_1}=2\phi$, $\phi^{(1)}_{B_0}=\pi-\phi$, and $\phi^{(1)}_{B_1}=\frac{\pi}{4}$. The effective relative angles between successive measurements are $\Delta^{A,a_1}_{x_1x_2}=\phi^{(2)}_{A_{x_2}|a_1}+\eta^A_{a_1|x_1}-\phi^{(1)}_{A_{x_1}}$, with an analogous definition for Bob. The corresponding angle matrices for the negative-outcome branches are $\Delta^{A,-}_{x_1x_2}=\Delta^{B,-}_{y_1y_2}=\begin{pmatrix}0&\pi\\-\pi&0\end{pmatrix}$, while those for the positive-outcome branches are $\Delta^{A,+}_{x_1x_2}=\begin{pmatrix}0&\frac{4\pi}{15}\\-\frac{4\pi}{15}&0\end{pmatrix}$ and $\Delta^{B,+}_{y_1y_2}=\begin{pmatrix}-\frac{\pi}{32}&-\frac{37\pi}{160}\\\frac{\pi}{9}&-\frac{4\pi}{45}\end{pmatrix}$. Rows and columns label the first- and second-round inputs, respectively. 
An explicit realization is provided in {\color{black}Sec.~IV-B of} the SM.


For each observer pair $(A_i,B_j)$, the CHSH correlator takes the form
\begin{align}\label{eq:triangle}
\mathcal{S}_{ij}
&=t^{(s)}_{ij}\sin 2\theta+t^{(c)}_{ij}\cos 2\theta+t^{(r)}_{ij}\nonumber\\
\end{align}
The coefficients depend only on the measurement parameters; their explicit expressions are provided in {\color{black}Sec.~IV B of }the SM.
All four conditions $\mathcal{S}_{ij}(\theta)>2$ hold on intervals with a nonempty intersection containing $[\pi/12,3\pi/32]$. Since $t^{(s)}_{ij},t^{(c)}_{ij}>0$ there, each correlator is strictly concave:
$\mathcal{S}_{ij}''(\theta)=-4t^{(s)}_{ij}\sin2\theta-4t^{(c)}_{ij}\cos2\theta<0$.
Hence, checking the endpoints suffices. We obtain $\min_{ij}\mathcal{S}_{ij}(\pi/12)\simeq2.0018$ and $\min_{ij}\mathcal{S}_{ij}(3\pi/32)\simeq2.0022$, establishing
\begin{equation}
\min_{i,j\in\{1,2\}}\mathcal{S}_{ij}(\theta)>2,
\quad
\forall\theta\in\left[\frac{\pi}{12},\frac{3\pi}{32}\right].
\end{equation}
Thus, a single fixed set of outcome-conditioned local instruments enables all four observer pairs to simultaneously violate the CHSH inequality over a finite interval of initial entanglement, analytically establishing fourfold sharing of Bell nonlocality under unbiased measurement-setting choices.

\begin{figure}[t]
	\centering
	\includegraphics[width=0.99\linewidth]{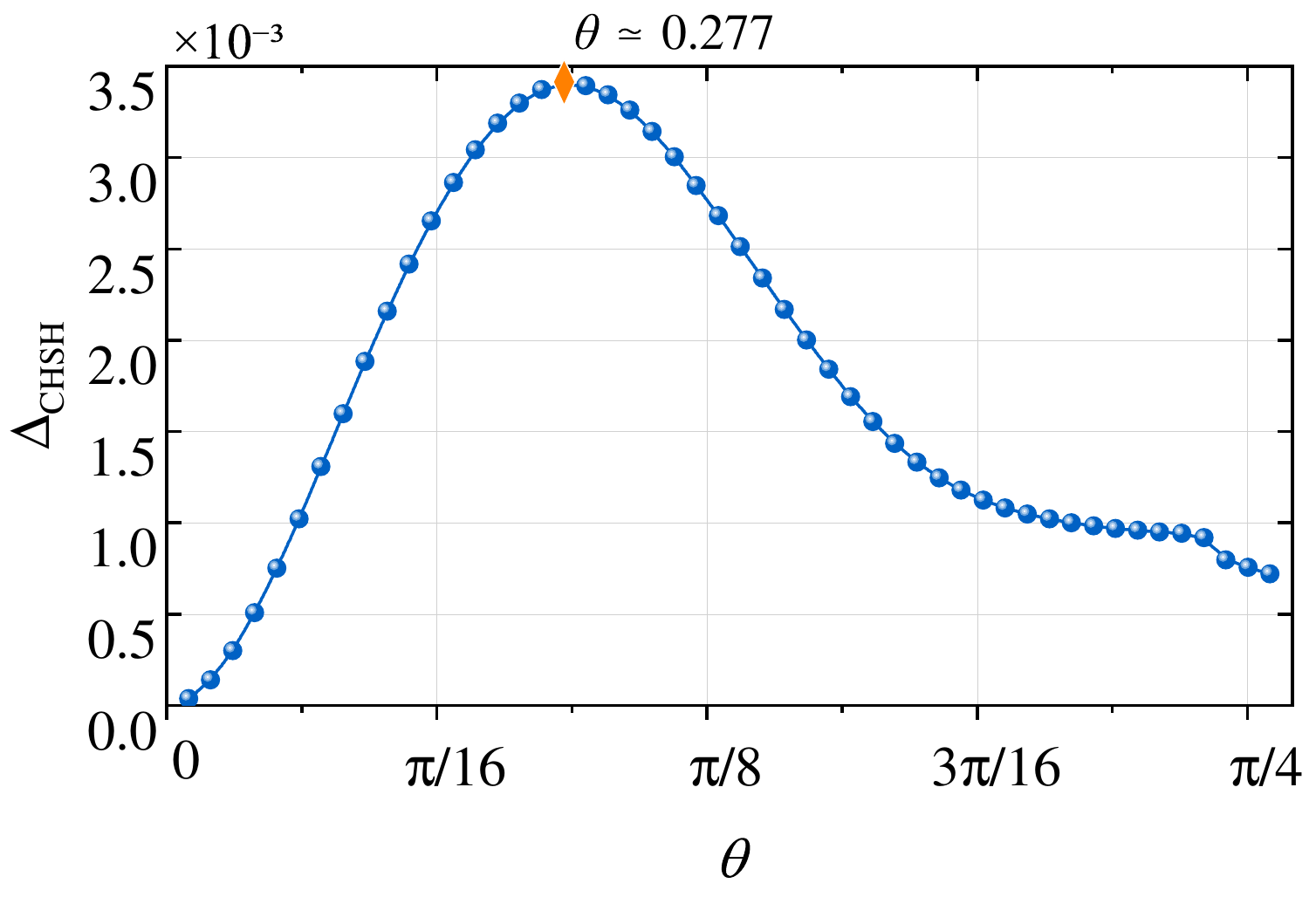}
	\caption{
Fourfold sharing of CHSH nonlocality. For each fixed initial-state parameter $\theta$, the local-instrument parameters are optimized to maximize the smallest CHSH value among the four observer pairs, yielding the common violation margin $\Delta_{\mathrm{CHSH}}(\theta)\equiv\mathcal V(\theta)-2$. The circles represent the optimized CHSH violation margins at 50 sampled values of $\theta$, connected by a line as a guide to the eye. Positive margins indicate that all four observer pairs simultaneously violate the CHSH inequality under the same local-instrument configuration. The highlighted point marks the largest sampled margin, $\Delta_{\mathrm{CHSH}}\simeq3.4\times10^{-3}$, at $\theta\simeq0.277$.
    }
	\label{fig:fourfold}
\end{figure}

We further numerically establish the existence of bilateral fourfold CHSH violation and determine its maximum achievable value. 
By optimizing the entanglement parameter and all local-instrument parameters defined in Eq. (1), we introduce
\begin{equation}
\mathcal V(\theta)
=
\max_{\boldsymbol{\lambda}\in\mathcal P}
\;
\min_{i,j\in\{1,2\}}
\mathcal{S}_{ij}(\theta;\boldsymbol{\lambda}),
\label{eq}
\end{equation}
where $\boldsymbol{\lambda}$ collects all tunable local-instrument parameters and $\mathcal P$ denotes their physical domain. We perform a max–min optimization over the full range $\theta \in [0,\pi/4]$, globally searching the local measurement parameters at each fixed value of $\theta$. Figure~\ref{fig:fourfold} presents the optimized results for 50 representative values of $\theta$. Under unbiased measurement settings, the maximum achievable common CHSH value is $V_{\max}\simeq 2.0034$, attained by a partially entangled state with $\theta\simeq 0.278$, rather than a maximally entangled state. 


To clarify the respective roles of Kraus-decomposition freedom and local feed-forward, we compare two restricted protocols: one excludes tunable local unitary operations while retaining outcome-dependent measurement settings, whereas the other removes outcome-dependent measurement settings while preserving the unitary freedom. Our numerical results, detailed in {\color{black}Sec.~IV\,C of} the SM, show that neither restricted protocol alone achieves fourfold CHSH violation. These findings highlight two complementary ingredients: Kraus-decomposition freedom, which enables outcome-conditioned local unitary feedback, and feed-forward of previous measurement outcomes to subsequent measurement settings. Their interplay enables bilateral fourfold sharing of Bell nonlocality.

\textit{Conclusions---} 
By developing a unified framework of quantum instruments, we characterized the sequential reuse of quantum correlations beyond simple measurement statistics. We showed that the efficacy of correlation sharing is governed by instrument-dependent backaction, which unifies seemingly different protocols into a single physical origin. Leveraging this framework, we demonstrated that bilateral Bell nonlocality can be shared even under unbiased measurements. Furthermore, we derived the rigorous conditions for infinite unilateral sharing, showing that nonlocality can be distributed across an arbitrary number of observers through tailored state and measurement designs. Our results indicate that the fundamental limits of correlation sharing are dictated by the internal structure of quantum instruments rather than mere POVMs, establishing a solid foundation for investigating quantum resource recycling in the future~\cite{resource_RevModPhys.91.025001}.

\appendix


\bibliography{ref.bib}

	
\end{document}